\documentclass[12pt,titlepage]{article}
\usepackage{setspace}
\usepackage[usenames]{color}
\usepackage{amsmath}
\usepackage{epstopdf}
\usepackage{graphicx}
\usepackage{textcomp}
\usepackage[round,numbers,sort&compress]{natbib}
\title{Coupled Reversible and Irreversible Bistable Switches Underlying TGF$\beta$-induced Epithelial to Mesenchymal Transition}

\author{Xiao-Jun Tian, Hang Zhang, and Jianhua Xing\thanks{Corresponding author. jxing@vt.edu} \\\\  Department of Biological Sciences, Virginia Tech,\\ Blacksburg, Virginia, 24061-0406, USA }

\date{}

\begin{document}

\maketitle

\abstract{Epithelial to mesenchymal transition (EMT) plays important roles in embryonic development, tissue regeneration and cancer metastasis. While several feedback loops have been shown to regulate EMT, it remains elusive how they coordinately modulate EMT response to TGF-$\beta$ treatment. We construct a mathematical model for the core regulatory network controlling TGF-$\beta$-induced EMT. Through deterministic analyses and stochastic simulations, we show that EMT is a sequential two-step program that an epithelial cell first transits to partial EMT then to the mesenchymal state, depending on the strength and duration of TGF-$\beta$ stimulation. Mechanistically the system is governed by coupled reversible and irreversible bistable switches. The SNAIL1/miR-34 double negative feedback loop is responsible for the reversible switch and regulates the initiation of EMT, while the ZEB/miR-200 feedback loop is accountable for the irreversible switch and controls the establishment of the mesenchymal state. Furthermore, an autocrine TGF-$\beta$/miR-200 feedback loop makes the second switch irreversible, modulating the maintenance of EMT. Such coupled bistable switches are robust to parameter variation and molecular noise. We provide a mechanistic explanation on multiple experimental observations. The model makes several explicit predictions on hysteretic dynamic behaviors, system response to pulsed stimulation and various perturbations, which can be straightforwardly tested.}

\vspace*{2cm}

\emph{Key Words: mathematical modeling, bifurcation diagram, cell differentiation, phenotypic transition,
gene regulatory network}
\clearpage

\section*{INTRODUCTION}

Epithelial to mesenchymal transition (EMT) is defined as the conversion of epithelial cell to mesenchymal cell, characterized by loss of cell-cell adhesion and increased cell motility \cite{Radisky2005, Nieto2011}. EMT plays important roles in embryonic development and tissue regeneration  \cite{Thiery2009, Nieto2011}. In addition, inappropriate utilization of EMT may lead to pathological processes, such as fibrosis and cancer metastasis \cite{Thiery2009}. Thus, in both physiological and pathological contexts understanding the regulation of EMT process is fundamental for prevention and treatment of cancer metastasis and organ-degenerative diseases.

EMT can be induced by a plethora of stimuli, of which transforming growth factor-$\beta$ (TGF-$\beta$) is a major and very potent inducer \cite{Heldin2009}.  Several crucial transcription factors are involved in TGF-$\beta$ mediated EMT, including SNAIL1/2, ZEB1/2. They directly repress the expression of the epithelial state (denoted as $E$ in this work) marker E-cadherin, while promote the induction of the mesenchymal state (denoted as $M$) marker N-cadherin \cite{Thiery2009}. Several miRNAs also are involved in regulating EMT, of which miR-34 and miR-200 are most studied \cite{Lamouille2013}. Moreover, the transcription factors and miRNAs are linked by several feedback loops. One is the double-negative feedback loop between ZEB1/2 (collectively referred to as ZEB thereafter) and miR-200 family (miR-200a, miR-200b, miR-200c, miR-141 and miR-429), in which ZEB1/2 represses the expression of miR-200 while miR-200 negatively regulates the translation of ZEB1/2 \cite{Wellner2009, Bracken2008, Burk2008}. The other double-negative feedback loop incurs between SNAIL and miR-34a/b/c (collectively referred to as miR-34 thereafter) in a similar way \cite{Siemens2011, Kim2011}. Furthermore,  release of miR-200-mediated inhibition of TGF-$\beta$ is required to maintain a stable mesenchymal phenotype \cite{Gregory2011}. That is, several coupled positive feedback loops seemingly function redundantly to regulate the process. It remains to be fully understood how these feedback loops coordinately modulate EMT response to TGF-$\beta$ treatment.

EMT is classically viewed as a switch that converts the static, ordered epithelia into labile, individual mesenchymal cells. However, EMT is not always an all-or-none response. In addition to the epithelial and mesenchymal states, reports exist on an intermediate phenotype in the process of EMT, known as partial EMT (referred as \textit{pEMT} thereafter) state \cite{Leroy2007, Revenu2009, Klymkowsky2009,  Thomson2011, Futterman2011}. The partial EMT state retains some characteristics of epithelium but also shows features of mesenchymal cells \cite{Herreros2010}. Notably, \textit{pEMT} is a metastable, reversible phenotype \cite{Lee2006}, while complete EMT is not. The existence of the three states, epithelia, \textit{pEMT}, and mesenchymal state, leads us to hypothesize that  complete EMT may be a sequential multistep program. Then the question is  how these three states are generated from the regulatory network and how partial EMT is reversible while full EMT is irreversible.

Several mathematical models have been developed to explore the dynamics of TGF-$\beta$ signaling pathway in epithelial mesenchymal transition \cite{Vilar2006, Zi2007, Schmierer2008, Turner2010}. In contrast to these two-compartment model, Vilar, Jose M. G. et al built a trafficking-coordinate model to accurately capture the distinct TGF-$\beta$ signaling dynamics \cite{Vilar2011}.  However, these models only focus on the quantitative dynamic of Smad nucleo-cytoplasmic shuttling and do not take into account the downstream of Smad signal, such as SNAIL1/miR-34 and ZEB/miR-200 double-negative feedback loops. Thus it remains to be further elucidated how the cell fate is coordinately determined during TGF-$\beta$-induced EMT.

Motivated by the above consideration, in this work we perform a systematic mathematical analysis on the core regulatory network for EMT, addressing the relation between the network structure and its functional roles. In the remaining parts we will first present the network constructed by integrating existing experimental studies, then discuss its deterministic and stochastic  dynamic behaviors in response to exogenous TGF-$\beta$ treatment and associate the network dynamics with cellular outcomes, and conclude with discussions on how the design of the network matches its functional requirements and with proposed future experimental studies for test our predictions from our simulation.

\section*{MODEL AND METHODS}

Figure 1 gives a schematic diagram of the core regulatory network. Autocrine or exogenous TGF-$\beta$ promotes the expression of \textit{snail1} mRNA \cite{Peinado2003},  which is translated to SNAIL1 protein. SNAIL1 promotes the transcription of \textit{zeb}, which is translated to ZEB. Experimental studies reveal two core double-negative feedback loops between transcription factors and miRNAs. The first one is between SNAIL1 and miR-34 \cite{Siemens2011, Kim2011}. SNAIL1 inhibits transcription of miR-34, which represses translation of \textit{snail1}, enclosing a double-negative feedback loop. The other one is between ZEB and miR-200  \cite{Wellner2009, Bracken2008, Burk2008}. ZEB represses  transcription of miR-200, which inhibits  translation of \textit{zeb}, enclosing the other double-negative feedback loop. miR-200 also inhibits the autocrine expression of TGF-$\beta$ \cite{Gregory2011}, making another feedback loop. As the marker of epithelial cells, E-cadherin is inhibited by SNAIL1 and ZEB, while the marker of mesenchymal cells N-cadherin is promoted by SNAIL1 and ZEB \cite{Thiery2009}. Table S1 in the Supporting Material summarizes the experimental support of the regulatory network. We first describe the dynamics of proteins, mRNAs and miRNAs with a set of  ordinary differential equations (ODEs), then study the effect of noises with stochastic simulations. We use the following set of ODEs to model the deterministic dynamics of the core regulatory network,

\begin{eqnarray}
\frac{d[\rm{T}]}{dt} &=& k0_{_{\rm{T}}}+\frac{k_{_{\rm{T}}}}{1 + (\frac{[\rm{R2}]}{J_{_{\rm{T}}}})^{n_{r2}}} - kd_{_{\rm{T}}}[T]\\
\frac{d[\rm{s}]}{dt} &=&  k0_{\rm{s}} + k_{\rm{s}} \frac{(\frac{[\rm{T}]+TGF0}{J_{\rm{s}}})^{n_{t}}}{ 1 + (\frac{[\rm{T}]+TGF0}{J_{\rm{s}}})^{n_{t}}}-kd_{\rm{s}}[\rm{s}]\\
\frac{d[\rm{\rm{S}}]}{dt} &=& k_{_{\rm{S}}}[\rm{s}]\frac{1}{ 1 + (\frac{[\rm{R3}]}{J_{_{\rm{S}}}})^{n_{r3}} }-kd_{_{\rm{S}}}[\rm{S}]\\
\frac{d[\rm{R3}]}{dt} &=& k0_{3} + k_{3}\frac{1}{1 + (\frac{[\rm{S}]}{J1_{3}})^{n_{s}} + (\frac{[\rm{Z}]}{J2_{3}})^{n_{z}}}-kd_{3}[\rm{R3}]\\
\frac{d[z]}{dt} &=& k0_{\textrm{z}} + k_{\textrm{z}}\frac{(\frac{[\rm{S}]}{J_{\textrm{z}}})^{n_{s}} }{ 1 + (\frac{[\rm{S}]}{J_{\textrm{z}}})^{n_{s}} }- kd_{\textrm{z}}[\textrm{z}]\\
\frac{d[\rm{Z}]}{dt}&=& k_{_{\rm{Z}}}[\rm{z}]\frac{1}{1 + (\frac{[\rm{R2}]}{J_{_{\rm{Z}}}})^{n_{r2}}} -kd_{_{\rm{Z}}}[\rm{Z}]\\
\frac{d[\rm{R2}]}{dt}&=&k0_{2} + k_{2}\frac{1}{ 1  + (\frac{[\rm{S}]}{J1_{2}})^{n_{s}} + (\frac{[\rm{Z}]}{J2_{2}})^{n_{z}}} -kd_{2}[\rm{R2}]\\
\frac{d[\textrm{E}]}{dt} &=& k_{\textrm{e1}} \frac{1}{(\frac{[\rm{S}]}{J1_{\rm{e}}})^{n_{s}} +1}+ k_{\textrm{e2}}\frac{1}{(\frac{[\rm{Z}]}{J2_{\rm{e}}})^{n_{z}}+1}-kd_{\rm{e}}[\textrm{E}]\\
\frac{d[\textrm{N}]}{dt} &=& k_{\textrm{n1}} \frac{(\frac{[\rm{S}]}{J1_{\rm{n}}})^{n_{s}}}{(\frac{[\rm{S}]}{J1_{\rm{n}}})^{n_{s}}+1}+ k_{\textrm{n2}}\frac{(\frac{[\rm{Z}]}{J2_{n}})^{n_{z}}}{(\frac{[\rm{Z}]}{J2_{\rm{n}}})^{n_{z}}+1}-kd_{\rm{n}}[\textrm{N}]
\end{eqnarray}

where [T], [s], [S], [R3], [z], [Z], [R2], [E] and [N] denote the concentrations of endogenous TGF-$\beta$, \textit{snail}, SNAIL, \textit{miR-34}, \textit{zeb}, ZEB, \textit{miR-200}, E-cadherin and N-cadherin, respectively. Regulation of the transcription of each gene is characterized by the Hill function, which is often used to represent the regulation of the transcription of an unknown gene \cite{Rosenfeld2005, Alon2007a}. Currently there is no experimental data on the response curve of the gene expression level versus the transcription factor level. For simplicity, we choose all the Hill coefficients to be 2 to allow sufficient nonlinearity. Tables  S2 and S3 list the definition and initial value of each variable and a set of standard parameter values respectively. The default state is the epithelial phenotype, in which the concentration of each specie is set to its stable steady state value in the absence of exogenous TGF-$\beta$ treatment (i.e., TGF0=0).

Intrinsic noise that results from low copy numbers of mRNAs and miRNAs per cell can affect all the biological processes \cite{Paulsson2004}. However, there are some systems that can function deterministically even when the average number of number of mRNA molecules is lower than one \cite{Vilar2002}. To investigate the stochastic effects of the finite number of molecules on the dynamics of EMT system, we also build a stochastic version of the model, which generally can be described by birth-and-death stochastic processes governed by chemical master equations. By introducing a system size factor $\Omega$, we change the concentration  of each molecular to its numbers, i.e., $ x_i= [x_i]\cdot\Omega$.  Table S4 lists all the reactions involved. We use the $\tau$-leap-based stochastic Gillespie algorithm to simulate the stochastic model \cite{marquez2007, cao2007}. To account for cell-to-cell variability in population level, we also add extrinsic noise in the model as static parameter value variations. In these single-cell based simulations,  every parameter value of a cell is randomly and uniformly chosen from 85\% to 115\% of its default value with Latin Hypercubic sampling.

\section*{RESULTS}

\subsection*{Epithelial to mesenchymal transition is a sequential two-step program}

Figure 2 shows the deterministic dynamics of proteins and miRNAs under different level of constant exogenous TGF-$\beta$ treatment. There are two possible cell fates in response to TGF-$\beta$ treatment. When the level of the exogenous TGF-$\beta$ is limited, e.g., 1.8 unit (ng/ml) (Fig. 2A), the mRNA of \textit{snail1}  is induced quickly. However, SNAIL1 protein level does not go up until the repression of translation mediated by miR-34 is removed later. Notably, the levels of ZEB and miR-200 barely change throughout the time. Furthermore, endogenous expression of TGF-$\beta$ is not activated. E-cadherin is partially inhibited and N-cadherin is partially activated.  That is,  the cell shows both epithelial and mesenchymal traits, consistent with the experimentally observed feature of partial EMT \cite{Herreros2010}.

When the level of the exogenous TGF-$\beta$ increases, e.g., 3.0 unit, the cell converts to a different cell type (see Fig.2B). First, the network dynamics undergoes a biphasic response. In the early stage, SNAIL1 is expressed and miR-34 is silenced. This induces partial inhibition of E-cadherin expression and partial activation of N-cadherin expression, and thus the cell reaches the \textit{pEMT} state. In this case the exogenous TGF-$\beta$ induces full expression of SNAIL1, which then induces transcription of \textit{zeb}. Again the increase of ZEB protein level is well delayed because of the translational repression mediated by miR-200. Eventual expression of both SNAIL1 and ZEB induces full inhibition of E-caderhin expression and full activation of N-caderhin expression, and finally shifts the cell to the mesenchymal state. Furthermore, the autocrine expression of TGF-$\beta$ is activated in the second phase after miR-200-mediated inhibition is released. The predicted biphasic dynamics is in good agreement with the experimental data, which is quantified from Ref.\cite{Bracken2008}. A more clear biphasic dynamics can be seen from the results with another human cell line MCF10A shown in Fig. S1, where more time course data points are available. Taken together, EMT is a sequential two-step programming, first to partial EMT and then to full EMT.

\subsection*{The two-step dynamics of EMT results from sequential activation of two bistable switches}

Close examination of the network in Fig. 1 suggests that the dynamics of EMT is governed by a cascade of two bistable switches, one shifts the cell to $pEMT$ state and the other triggers the cell to full EMT. This mechanism can be clearly seen from the one-parameter bifurcation analysis in Fig. 3A, which shows  dependence of the steady states of N-cadherin on the exogenous TGF-$\beta$ level. There are three distinct branches (solid lines) in the bifurcation diagram, indicating existence of three types of stable steady states: $E$ (blue lower branch), $pEMT$ (green middle branch) and $M$ (red upper branch) states. These three states are governed by two bistable switches. Suppose that one cell originally starts  in the lower branch corresponding to the $E$ state. Upon gradually adding exogenous TGF-$\beta$, the N-cadherin level jumps to the middle branch after it crosses the first saddle-node bifurcation point SN1. That is, the cell is transited to the $pEMT$ state from the $E$ tate. While at the $pEMT$ state,  if the exogenous TGF-$\beta$ level decreases beyond another saddle-node bifurcation point (SN2), the N-cadherin level shifts back to low level corresponding to the $E$ state. The exogenous TGF-$\beta$ level at SN1 is larger than that at SN2 and the latter is larger than zero. That is, the transition from $E$ to $pEMT$ state is governed by one reversible bistable switch.

When the level of exogenous TGF-$\beta$ is larger than that at SN3,  the N-cadherin level is shifted to the upper branch from the middle branch, i.e., the cell is transited to the $M$ state from the $pEMT$ state. While at the $M$ state, the cell stays on this state even if the exogenous TGF-$\beta$ is removed. Actually, there is another saddle-node bifurcation point SN4 on the negative side of the axis, which can not be reached by the system physically. That is, the transition from the $pEMT$ state to the $M$ state is governed by one irreversible bistable switch.

Figure 3B-D gives a full spectrum of the system dynamics under pulsed exogenous TGF-$\beta$ with different strength and duration. The maximum level of N-cadherin shows three regions in the plane of strength and duration of exogenous TGF-$\beta$. Treatment of exogenous TGF-$\beta$ in the light gray (yellow online) region triggers the cell to $pEMT$. However, the cell recovers back to $E$ state after the stimulus is withdrawn since the endpoint level of N-cadherin recovers to low level, showing the reversibility of partial EMT. In contrast, the treatment of exogenous TGF-$\beta$ in the dark gray (red online) region triggers the cell to full EMT and maintains it in the $M$ state after the stimulus is withdrawn, exhibiting the irreversibility of the full EMT. Therefore, complete EMT requires both a minimal duration and a minimal strength of TGF-$\beta$ treatment.

To see whether or not this conclusion depends heavily on parameter values, we perform parameter sensitivity analysis to systematically explore how the features of two switches change upon varying each of the 33 parameters in the feedback loops  and the Hill coefficients. The variation of the parameters in the expression of E-cadherin and N-cadherin do not change four thresholds. The thresholds for activation of partial EMT and full EMT are the corresponding exogenous TGF-$\beta$ levels at SN1 and SN3 respectively.  The reversibility/irreversibility of partial/full EMT is determined by the sign of the corresponding exogenous TGF-$\beta$ level at SN2/SN4. First from Fig. 4A, all the four saddle-node points are distributed around that under the standard parameter set when each parameter is increased or decreased by 15\% with respect to the standard parameter set. Notedly, one sees that the corresponding exogenous TGF-$\beta$ level at SN2 is always positive while that at SN4 is always negative under this parameter variation.  That is, the first switch is robustly reversible while the second one is robustly irreversible with this degree of parameter variation.

We then further calculate the percentage of changes in four thresholds in Fig. 4B. The thresholds for the first switch are sensitive to the parameters in the SNAIL1/miR-34 feedback loop, while the thresholds for second switch are sensitive to the parameters in both the SNAIL1/miR-34 and ZEB/miR-200 feedback loops. Furthermore, SN4 is also sensitive to the parameters associated with the autocrine expression of TGF-$\beta$, implying the regulation of autocrine TGF-$\beta$ on the irreversibility of the second switch. We will come back to this point later. In summary, the EMT network has a robust basic feature of a reversible bistable switch coupled to an irreversible bistable switch.

\subsection*{Coupled positive feedback loops are responsible for the two bistable switches}

The above analysis demonstrates that the dynamics of EMT is governed by two bistable switches. To reveal the underlying mechanism for these two bistable switches, we examine the dependence of steady-state levels of  SNAIL1, miR-34, ZEB, miR-200 and the endogenous TGF-$\beta$ on the exogenous TGF-$\beta$ level, as shown in Fig. 5A. When the exogenous TGF-$\beta$ level is less than that at SN1, the expression level of SNAIL1 is low, and that of miR-34 is high. They switch their dominance at SN1. The expression level of SNAIL1 change dramatically again at SN3. The miR-34 levels for the $pEMT$ and $M$ states are both very low and hardly distinguishable. However, experimentally if one starts to reduce the  exogenous TGF-$\beta$ level, the system initially at\textit{pEMT} state can eventually jump back to the high miR-34 level \textit{E} state. On the other hand, the system initially at \textit{M} state will maintain a low miR-34 level.

In contrast, only the two bifurcation points at SN3 and SN4 (in the negative side of axis) are easily detectable in the signal response curves of ZEB, miR-200 and autocrine TGF-$\beta$.  When the exogenous TGF-$\beta$ level is larger than that at SN3, ZEB is expressed while miR-200 is inhibited. That is,  ZEB and miR-200 dominate alternatively with the second switch on and off, respectively. Thus, the second irreversible bistable switch depends on the ZEB/miR-200 feedback loop. Turning on the second switch also activates the autocrine expression of TGF-$\beta$, which amplifies the level of SNAIL1 to further level, and leads to the four clear four bifurcation points on the response curve of SNAIL1.

It is well known that positive feedback loop is a basic motif to function as a bistable switch \cite{Tyson2003, Ferrell2002, Yao2008, Tian2009}. To further investigate the correlation between two bistable switches and SNAIL1/miR-34 and ZEB/miR-200 double negative feedback loops, Fig. 5B shows the dependence of the two thresholds for activation of partial EMT and full EMT on the feedback strength of two loops. For simplicity, the Michaelis constant of mutual inhibition between SNAIL1 and miR-34 ($J_{_{\textrm{SNAIL}}}$, $J_{{\textrm{34}}}$) are chosen to characterize the feedback strength of the first loop, while the Michaelis constant of mutual inhibition between ZEB and miR-200 ($J_{_{\textrm{ZEB}}}$, $J_{{\textrm{200}}}$) are chosen to characterize the feedback strength of the second loop. Both two thresholds decrease with the increase of $J_{_{\textrm{SNAIL}}}$ while increase with $J_{\textrm{34}}$. That is, both induction of partial EMT and full EMT can be regulated by SNAIL1/miR-34 feedback loop. By contrast, only the threshold for full EMT induction is regulated by the feedback strength of the second loop. Thus, EMT is initiated by Snail/miR-34 regulatory loop, and stabilized by ZEB/miR-200 regulatory loop. This is consistent with experimental report \cite{Gregory2011}. Furthermore, activation of the second switch depends on activation of the first switch. Taken together, the two-step dynamics of EMT results from sequential activation of two bistable switches, which are governed by SNAIL/miR-34 and ZEB/miR-200 double negative feedback loops.

\subsection*{Manipulation of the autocrine TGF-$\beta$ modifies the irreversibility of the full EMT}

The results discussed above show that the autocrine expression of TGF-$\beta$ is activated when the second bistable switch is turned on. To further investigate the role of autocrine expression of TGF-$\beta$ on the dynamics of EMT, we perform \textit{in silico} experiments of adding a constant TGF-$\beta$ expression inhibitor or  miR-200 expression activator. Figure 6A shows that the thresholds SN1$\sim$3 barely change with the inhibitor level. On the other hand, the threshold for the deactivated of the second bistable switch (SN4) changes from negative to positive with increasing the inhibitor level. The same goes for the dependence of SN4 on miR-200 activator level (Fig.S2), a alternative way to perturbing TGF-$\beta$ expression. That is, the second bistable switch is turned to be reversible when the TGF-$\beta$ inhibition or miR-200 activation is large enough. The bifurcation diagrams shown in Fig. 6B and Fig. S2B clearly shows this transition from an irreversible (wild type) to a reversible (mutant) switch.

Thus, the irreversibility of the second bistable switch is only relative. It can be tuned by modulating signals in addition to the exogenous TGF-$\beta$, explaining the stable yet reversible feature of EMT. Taken together, release of the inhibition of miR-200 on autocrine expression of TGF-$\beta$ is responsible for the irreversibility of the full EMT induction. This is consistent with experimental results that maintenance of a stable mesenchymal phenotype requires the establishment of autocrine TGF-$\beta$ signaling \cite{Gregory2011}. It should be noted that the irreversibility of the full EMT is cell-type-dependent. It has been shown that the transition to mesenchymal state is stabilized after prolonged exogenous TGF-$\beta$ treatment in MDCK cells following withdrawal of the TGF-$\beta$ \cite{Gregory2011}, but it can not be sustained long term after TGF-$\beta$ withdrawn in epithelial NMuMG cells \cite{Gal2008}. The reason for this difference is presently not well understood and needs further study. Our results suggest one possibility is either reduced expression of autocrine TGF-$\beta$  or overexpression of miR-200 in this cell line compared to others.

The requirement of the autocrine expression of TGF-$\beta$ on the irreversibility of EMT indicates that the conversion from mesenchymal to epithelial (MET) needs suppression of endogenous TGF-$\beta$ signaling by exogenous factors to prevent  self-perpetuation of EMT. Actually, during the initiation phase of reprogramming of mouse fibroblasts, exogenous factors OSKM (Oct4, Sox2, Klf4, Myc) suppress TGF-$\beta$ signaling \cite{Li2010}, both Oct4/Sox2 and BMP can directly activate miR-200 to induce MET \cite{Samavarchi2010, Wang2013}. These results are consistent with our finding here that it needs to turn the second bistable switch to be reversible for induction of MET by inhibiting the autocrine expression of TGF-$\beta$. Interesting,  partial MET state exists in the transition from the M state to the EMT state (Fig.S3). Consistently, there is one experimental report on the partial MET state \cite{Chao2012}.

\subsection*{Influences of the intrinsic and extrinsic noise on the epithelial to mesenchymal transition}
Cells face various kinds of uncertainties, such as intrinsic noises and extrinsic variations.  The intrinsic noises come from  low numbers of molecules per cell, such as  miRNAs in our model, while the extrinsic noises result from parameter variations due to fluctuations of the environment including other cellular components. They generate phenotypic heterogeneity at the population level.

Figure 7A shows 50 typical simulated stochastic trajectories of cells under 2 units of exogenous TGF-$\beta$. The  biphasic dynamics survives over inclusion of noises,  although the level of N-cadherin in $pEMT$ state and $M$ state varies from cell to cell. It is noticed that the backward transition from the pEMT state to the E state in Fig.7 A is very rare. This asymmetry of forward and backward transition rates is related to hysteresis, a  prominent property of   a bistable system. That is, the required TGF-$\beta$ concentration to induce the forward transition is higher tha that to induce the backward transition. While we choose a TGF-$\beta$ concentration that favors forward transition, the backward transition rate is very low under this level of noise. Indeed, if we consider that the exogenous concentration of TGF-$\beta$ fluctuates (so it can be close to SN1 or SN2 at different time), as shown in Fig.S5, one can observe more frequent forward/backward transitions between the pEMT state and the E state.

More prominent cell-to-cell variation is the time a cell transits to $pEMT$, denoted as $T_{pEMT}$, and the time of transition to $M$ state, $T_{EMT}$.  Both the distributions of $T_{pEMT}$ and $T_{EMT}$, especially that of $T_{EMT}$, show broad distributions either when incorporating both extrinsic and intrinsic noise (Fig. 7B) or only intrinsic noise (Fig. S4). This is consistent with the sensitivity analysis in Fig.4, the threshold for activation of full EMT (SN3) is more sensitive to parameter variation than the threshold for activation of pEMT (SN1).

The time evolution of each subpopulation in Fig. 7C shows typical three-state dynamics under 2 units of TGF-$\beta$. Starting with cells in the epithelial state, the $E$ state cell percentage drops monotonically, the $pEMT$ fraction firstly increases then degreases, leading to  increase of $M$ state cells. The $pEMT$ and $M$ state cells coexist after a long time since the TGF-$\beta$ concentration is within the bistable region. We notice that even at the end of the simulation, 20 days, a significant portion of cells still remain in the $E$ state, which is beyond the expectation from a simple exponential decay dynamics. The evolution over time is more clearly seen from the snapshots in Fig. 7D. The 2000 simulated cells initially have a narrow distribution centered at low N-cadherin levels, then gradually another peak with medium N-cadherin level appears and grows to dominate over the low N-cadherin peak, followed by appearance of the third peak with high N-cadherin levels. One can experimentally test on these dynamic behaviors using flow cytometry measurements.

\section*{DISCUSSION}

\subsection*{The core regulatory network is  designed to be robust against fluctuations and response differentially to signals with different strength and duration}
Epithelial cells form the  densely packed top layer of cavities, glands, and other structures throughout the body. Each individual cell adheres to its neighbors tightly and is highly immobile. On the other hand, mesenchymal cells are loosely connected to each other, and thus easy to move. Conversion between these two distinct phenotypes requires global change of the transcription program \cite{Thiery2009}, and is expected to be tightly regulated to prevent inadvertent transitions. At certain physiological conditions, such as wound healing, however, some epithelial cells need to response to certain stimuli, migrate to the target site, and then convert back to epithelial cells after the stimulus has been removed, e.g., after the wound healing process is finished. Therefore the regulatory network has the combined functional requirements of being both robust against stochastic fluctuations and  flexible to convert between the immobile and migratory phenotypes reversibly.

Our analyses in this work suggest that the EMT core regulatory network has evolved in several aspects to achieve the above requirements. The basic structure of the network is a cascade of  two self-activation motifs. Each of the motifs serves as a bistable switch, with the end product of the first one, SNAIL1, serves as the trigger for the second switch. This allows the existence of an intermediate state, the partial EMT state, with only the first switch turned on, at intermediate stimulus strength and duration, as shown in Fig. 2A and Fig. 3.  Figure 8 summarizes the overall picture of the whole process. Upon removal of the external stimulus, this partial EMT state switches back to the epithelial state since no mechanism exists to sustain the state. The two switches are further connected by a miR-200/TGF-$\beta$ mediated positive feedback. Consequently, with sufficiently long and strong exogenous stimulus to turn on the second switch, TGF-$\beta$ can be produced endogenously, allowing the switch to remain at the "on" state even after removing exogenous TGF-$\beta$.

Each switch is composed of a double negative feedback loop mediated with miRNAs. This network motif is known for generating delayed responses \cite{Alon2007}. The delay is further facilitated by the long time required to remove the inhibiting miRNAs, which have very long half life and their concentrations are mainly reduced by cell-division related dilution. The delayed responses serve as low-pass filters to prevent unregulated cell phenotypic transitions due to transient  TGF-$\beta$ concentration fluctuations within the  \textit{in vivo} microenvironment surrounding a cell.

Due to intrinsic and extrinsic noises, we predict that EMT shows stochastic transition dynamics with a broad distribution of the transition time and varying sensitivity to TGF-$\beta$. The population relaxation in Fig. 7C is slower than an exponential decay process, which is analogous to the "dynamics disorder" phenomenon observed in macromolecule dynamics \cite{Zwanzig1990, English2005}. This heterogeneous response within a population may have evolutionary advantage for generating an even broader response spectrum and preventing exhausting the whole population with a single stimulation.

\subsection*{Our model analyses make several experimentally testable predictions}
While seemingly simple, the regulatory network shown in Fig. 1 gives rise to complex dynamic features. Theoretical analysis as done in this work provides essential guidance for experimental studies on the structure-function relations of the network. Table 1 summarizes the published experimental reports that are qualitatively consistent with our simulation results, but more systematic quantitative measurements are necessary. Furthermore several important new testable predictions await for confirmation. Below we suggest a list of experimental studies.

First,  in Fig. 2 we predict that  at intermediate exogenous TGF-$\beta$ concentration, concentrations of the molecular species associated with the first switch, but not those of the second switch, change significantly, while at high exogenous TGF-$\beta$ concentration, the two switches turn on in sequence. Current existing studies only report the time course of E-cadherin or N-cadherin, and the data points are too sparse to unambiguously tell the predicted biphasic dynamics. Therefore, detailed quantitative time course and steady state measurements  on multiple involved molecular species are necessary for characterizing the system dynamics and unraveling its mechanism.

Second, the bifurcation diagrams shown in Fig. 3A and  Fig. 5A, and the result in Fig. 7 can be tested using some standard procedures. Through slowly varying the exogenous TGF-$\beta$ concentration as indicated in Fig. 3A, the model predicts that the N-cadherin level, which can be measured by methods such as western blot, shows hysteresis, \textit{i.e.}, at a given exogenous TGF-$\beta$ concentration within a bistable region, the N-cadherin level is different depending on whether the cells are originally prepared with low or high exogenous TGF-$\beta$ concentrations. This procedure follows what used in the classical studies of Sha \textit{et al.} \cite{Sha2003} and Pomerening \textit{et al.} \cite{Pomerening2003} to confirm the bistable cell-cycle transitions in Xenopus laevis egg extracts predicted by the Tyson-Novak model \cite{Novak1993}. The hysteretic dynamics predicted in Fig. 5A can be tested similarly.  More directly, the multimodal distributions predicted in Fig. 7 can be tested with flow cytometry, which have been used in other studies demonstrating multistability of cellular systems \cite{Yao2008}.  Microfluidic devices in combination with some fluorescence labeling may facilitate systematic measurements under different conditions, such as TGF-$\beta$ stimulation with different strength and duration predicted in Fig. 3B.  Further molecular interference, such as siRNA may test on how varying strengths of different interactions affects the reversibility of each switch, as predicted in Fig. 5B and Fig. 6B. The new experimental data can then be fed into the mathematical model for further analysis.

Third, our numerical analysis suggests possible missing regulation mechanism for miRNAs. Studies show that miRNAs  are very stable and have bare half life about 120 hours \cite{Gantier2011}. Therefore change of the cellular concentration of miRNAs is mainly due to cell division and grow related dilution, which gives an effective half life of one cell cycle time. We use this half-life time in all the results reported in the main text. However,  the MCF10A data shown in Fig. S1 reveals a much faster dynamics. We have to use a smaller half-life of miRNAs, $\sim 5$ hours, to reproduce the data. We thus suggest the dynamics of miRNAs may be regulated by additional mechanisms not reported for this system. Possible mechanisms include some post-transcriptional regulated degradation of miRNAs \cite{Siomi2010} or competitive inhibition of miRNAs \cite{Ebert2007}. Alternatively, another pathway other than miRNAs may exist to regulate SNAIL1 and E-cadherin expression and responses to TGF-$\beta$. Explicit time course measurements of the miRNAs, as shown in Fig. S1,  will help on clarifying this seemingly controversy.

\subsection*{Several directions exist on future computational/experimental studies}

The regulation of EMT is controlled in several layers including the transcriptional, translational machinery, post-translational regulation, expression of non-coding RNAs, stability and subcellular localization \cite{Craene2013}. Recent evidences also show that epigenetic changes across the genome also occur during EMT \cite{McDonald2011}, and prolonged treatment of TGF-$\beta$ leads to DNA hypermethylation of the miR-200 promoters \cite{Gregory2011}. Here, we only focus on the transcriptional, translational, non-coding RNAs layers, and leave inclusion of other layers for future studies.

In this work we only consider the core regulatory network, and other regulators in EMT are not treated here, such as Smad \cite{Schmierer2008}, Twist\cite{Dave2011} and other microRNAs \cite{Lamouille2013}. These regulators are either the upstream of SNAIL1/ZEB or more cell dependent than that of SNAIL1 and ZEB.  Further more, several other signaling pathways, such as Wnt, Notch and Ras-MAPK, can cooperate with TGF-$\beta$ to induce EMT \cite{Thiery2009}. It is an interesting and important question how these factors and pathways work cooperatively to regulate the EMT dynamics.

When endogenous TGF-$\beta$ molecules are induced and secreted out of cells, the extracellular TGF-$\beta$ in the local cellular environment increases, inducing EMT of more cells. That is, there is some auto-activation effect among cells at the tissue level. In this work we assume that TGF-$\beta$ diffusion is fast at the experimental culturing conditions, and thus the extracellular TGF-$\beta$ concentration is uniform. At the tissue level where cells are densely packed in three-dimensions, some TGF-$\beta$ gradient may be established. It is interesting to investigate this effect and physiological consequences by build a spatial model with cell-cell communication together with related experimental studies in the future.

Reports exist that the miR-200 and miR-34 are targets of the tumor suppressor p53 \cite{Hermeking2012}. That is, both ZEB/miR-200 and SNAIL1/miR-34 double negative feedback loops are regulated by p53. Thus, besides the conventional roles of regulating cell cycle arrest, senescence and apoptosis, p53 has a regulating role in controlling EMT-MET plasticity. Since inappropriate occurrence of EMT leads to cancer metastasis \cite{Chaffer2011,Craene2013}, it is interesting to explore the role of p53 in cancer metastasis through regulating EMT. It also reported that EMT is regulated by the PI3K/PTEN/AKT pathway through Twist and Snail \cite{Hao2012}, and  PI3K/PTEN/AKT pathway also plays important role in cell fate decision when coupled with p53 signalling pathway \cite{Zhang2011, Tian2012}. And various reports show that ZEB family also link EMT with cell cycle control, apoptosis and senescence \cite{Browne2010}. Thus, another important question to be answered is how cell fates are decided among EMT, apoptosis and cell cycle arrest.

The reverse process of EMT, MET, is a key step in cell reprogramming. It has been shown that MET is the first step during the reprogramming of mouse fibroblasts into iPSCs by exogenous factors OSKM (Oct4, Sox2, Klf4, Myc), which suppress TGF-$\beta$ signaling \cite{Li2010}. It also has been shown that BMP dependent activation of miR-200 cooperate with OSKM drives MET during the initiation phase of reprogramming of mouse fibroblasts \cite{Samavarchi2010}. Recently, it is found that Oct4/Sox2 can also directly activate  miR-200 \cite{Wang2013}.  As the reverse process of MET, the mechanism of EMT presented here may provide clues for improving the efficiency of cell reprogramming. Furthermore, EMT is associated with stemness properties and is potentially involved in cancer stem cells, since that stem cell factors, such as Sox2 and Klf4 are also candidate targets of miR-200 family \cite{Wellner2009}. Understanding the underlying mechanism will give some ideas of effective therapeutic strategy to prevent tumor metastasis.

\section*{SUPPORTING MATERIAL}

Supporting methods, figures, and tables are available at ...\\

\hskip -1em We thank Dr John Tyson for helpful discussions.\\

\hskip -1em This work has been supported by National Science Foundation Grants DMS-0969417, DGE-0966125, and NIH grant AI099120.

\section*{SUPPORTING CITATIONS}
References  \cite{Brabletz2010, Brabletz2012, Braun2010, Wang2011, Cano2000, Batlle2000, Shirakihara2007,  Moreno-Bueno2006, Zhou2004,Vandewalle2005, Sharova2009, Schwanhausser2011} appear in the Supporting Material.


\clearpage

\begin{table*}\small
\caption{\bf{Summary of predictions and  experimental support of simulation results}}\vspace{8pt}
\begin{tabular}{p{0.9\textwidth}|p{0.12\textwidth}}
\hline
Descriptions&Ref.\\
\hline

EMT is a sequential two-step programm, there are two distinct phase in EMT (initiation and maintenance). & \cite{Tran2011}\\

Partial EMT retains incomplete decrease of the epithelial marker also shows features of mesenchymal marker. & \cite{Ansieau2008}\\

Short-term TGF-$\beta$ treatment causess a partial, reversible EMT (\textit{pEMT}).  & \cite{Jechlinger2006, Waerner2006, Gregory2011}\\

Complete EMT requires a minimal strength and duration of TGF-$\beta$ treatment.
&\cite{Gregory2011}\\

SNAIL1 is uniquely required for the initiation of TGF-$\beta$-induced EMT and an essential early mediator of activation of the ZEB/miR-200 pathway. & \cite{Dave2011, Tran2011}\\

TGF-$\beta$ induces EMT via upregulation of SNAIL1 at early states of the transition while subsequent expression of ZEB  maintains the mesenchymal, migratory phenotype.& \cite{Dave2011, Thuault2008}\\

SNAIL1 initiates the repression of E-caderhin expression, ZEB is required for the completer inhibition.& \cite{Peinado2007, Herreros2010}\\

Two-phasic behavior is governed by a cascade of one reversible and  one irreversible bistable switches. & {Prediction}\\

The SNAIL1/miR-34 double negative feedback loop is responsible to the first bistable reversible switch and triggers the initiation of EMT. & {Prediction}\\

The ZEB/miR-200 double negative feedback loop is accountable to the second bistable irreversible switch. & {Prediction} \\

Release of the miR-200-mediated inhibition on autocrine TGF-$\beta$ makes the complete EMT irreversible &\cite{Gregory2011}\\

\hline
\end{tabular}
\end{table*}

\clearpage
\section*{Figure Legends}

\subsubsection*{Figure~1}
Schematic depiction of the core regulatory network. Exogenous or endogenously expressed TGF-$\beta$ promotes the expression of \textit{snail1} mRNA, which is translated to SNAIL1. SNAIL1 promotes the transcription of \textit{zeb} mRNA, which is translated to ZEB. SNAIL1 inhibits the expression of miR-34, which represses the translation of \textit{snail1}, forming a double-negative feedback loop. ZEB represses  the induction of miR-200, which inhibits the translation of \textit{zeb}, forming another double-negative feedback loop. The autocrine expression of TGF-$\beta$ is also inhibited by miR-200. As the marker of epithelial cell, E-cadherin is inhibited by SNAIL1 and ZEB, while as the marker of mesenchymal cell, N-cadherin is promoted by SNAIL1 and ZEB.

\subsubsection*{Figure~2}
Epithelial to mesenchymal transition is a sequential two-step program. (A) The dynamics of partial EMT under 1.8 unit treat of exogenous TGF-$\beta$. (B) The predicted (solid lines) and experimentally measured (circles) dynamics of full EMT under 3.0 unit treat of exogenous TGF-$\beta$. The experimental data on MDCK cell line is quantified from Ref.\cite{Bracken2008}.

\subsubsection*{Figure~3}
The biphasic EMT dynamics results from two coupled bistable switches. (A) Bifurcation analysis of N-cadherin level versus the exogenous TGF-$\beta$ level. (B-D) Phase diagrams showing N-cadherin maximum and end-point values for cells subjecting to pulse treatment of exogenous TGF-$\beta$ with different strength and duration.

\subsubsection*{Figure~4}
Parameter sensitivity of the two bistable switches. (A) The level of exogenous treatment of TGF-$\beta$ at the four saddle-node bifurcation points SN1$\sim$4 when each of 33 parameters in the coupled feedback loops and the Hill coefficients is increased or decreased by 15\% with respect to the standard parameter set. (B) Percentage of changes in the levels of exogenous TGF-$\beta$ at SN1$\sim$4 compared to that with the standard parameter set.

\subsubsection*{Figure~5}
Bifurcation analysis indicates that EMT is initiated by Snail1/miR-34 regulatory loop, and stabilized by ZEB/miR-200 regulatory loop. (A) Bifurcation diagrams of  SNAIL1, miR-34, ZEB, miR-200, and endogenous TGF-$\beta$ on the level of exogenous TGF-$\beta$; (B) Dependencies of the thresholds for activation of partial EMT and full EMT on the strength of SNAIL1/miR-34 feedback loop ($J_{_{\rm SNAIL}}$ and $J_{\rm 34}$ ) and the strength of ZEB/miR-200 feedback loop ($J_{_{\rm ZEB}}$ and $J_{\rm 200}$).

\subsubsection*{Figure~6}

Irreversibility of full EMT is maintained by the endogenous  TGF-$\beta$ expression. (A) Dependence of four bifurcation points on the level of endogenous TGF-$\beta$ expression inhibitor. (B) Bifurcation diagrams of N-cadherin depending on the exogenous level of TGF-$\beta$ for wild type cells (Curves) and mutant cells (Curves with circles) with inhibited endogenous expression of TGF-$\beta$.

\subsubsection*{Figure~7}

Influence of the intrinsic and extrinsic noises on the epithelial to mesenchymal transition. (A) N-cadherin dynamics of 50 cells  from stochastic stimulation. (B) Distribution of the time for transition to partial EMT $T_{_{pEMT}}$ and that for transition to mesenchymal state  $T_{_{\rm EMT}}$. (C) The percentage of cell fates as a function of time. (D) N-cadherin distribution at different time points. The exogenous level of TGF-$\beta$ treatment is 2.0.
\subsubsection*{Figure~8}
Schematic depiction of the mechanism of EMT. In epithelial cells,  expression of SNAIL1 and ZEB are inhibited by miR-34 and miR-200 respectively. Under treatment of exogenous TGF-$\beta$, cells first transit into a partial EMT state by the first reversible bistable switch, which is governed by the SNAIL1/miR-34 double negative feedback loop. Then if the strength and duration of treatment are large enough, cells transit to the mesenchymal state by the second irreversible bistable switch, which is governed by the ZEB/miR-200 and autocrine expression of TGF-$\beta$. For each molecular specie (node), a dark gray box denotes high expression level and an light gray box refers to low expression level; while the solid lines refer to activated regulation and the dashed lines indicate the inactivated regulation.

\begin{figure}
  \begin{center}
    \includegraphics[width=5in]{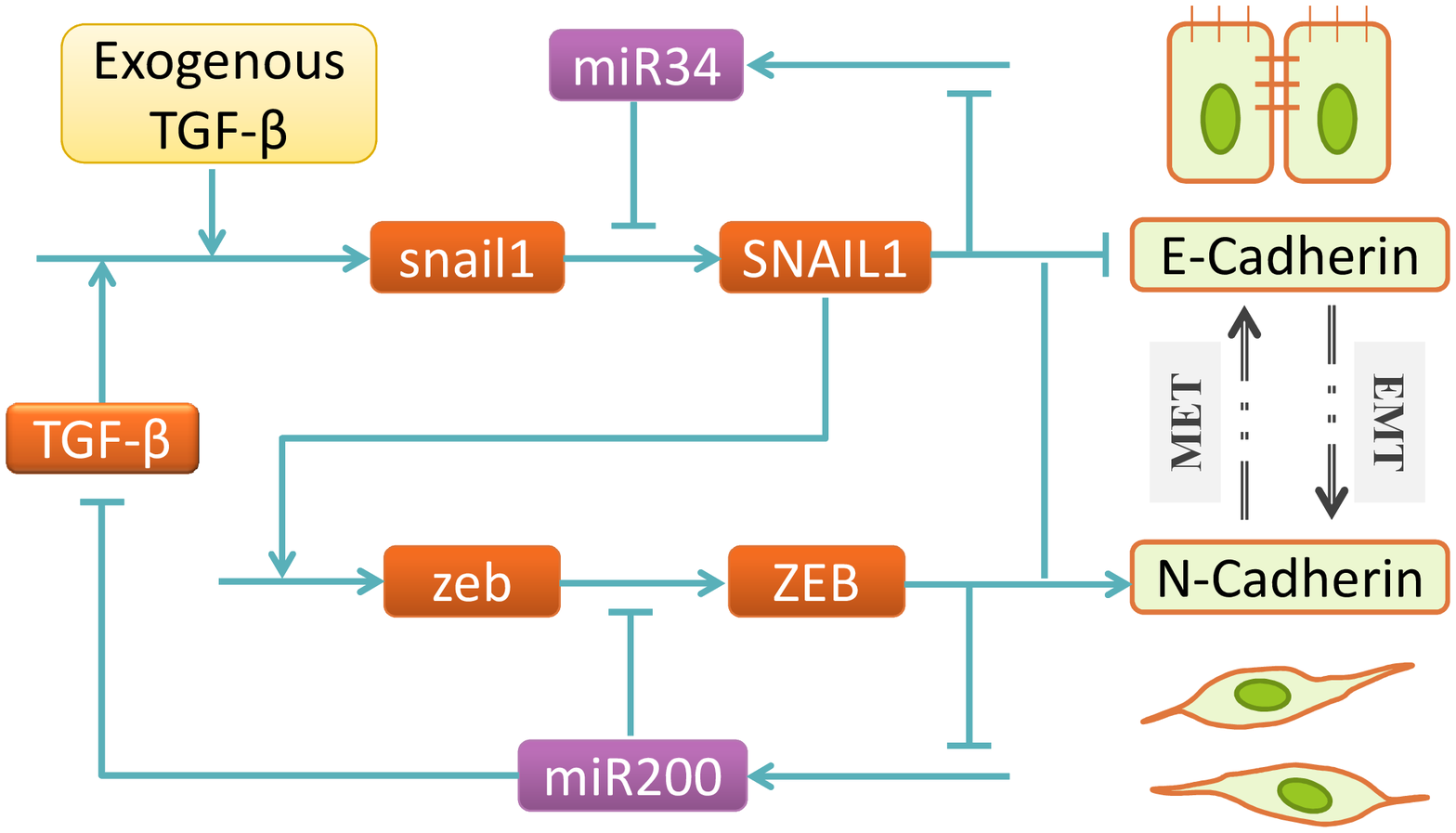}
    \caption{}
    \label{fig:one}
  \end{center}
\end{figure}

\clearpage
\begin{figure}
  \begin{center}
    \includegraphics[width=5in]{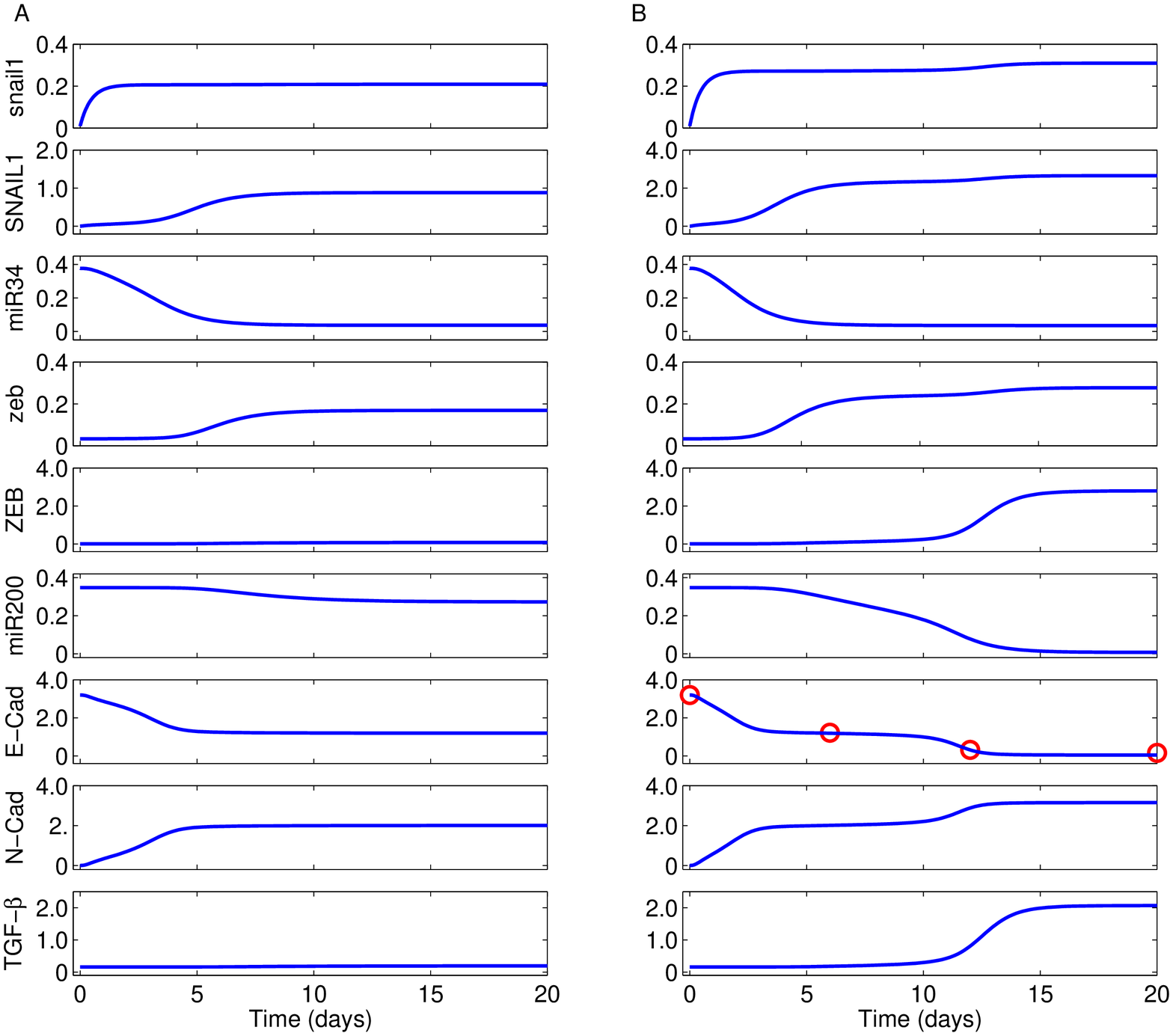}\\
    \caption{}
    \label{fig:two}
  \end{center}
\end{figure}

\clearpage
\begin{figure}
  \begin{center}
    \includegraphics[width=5in]{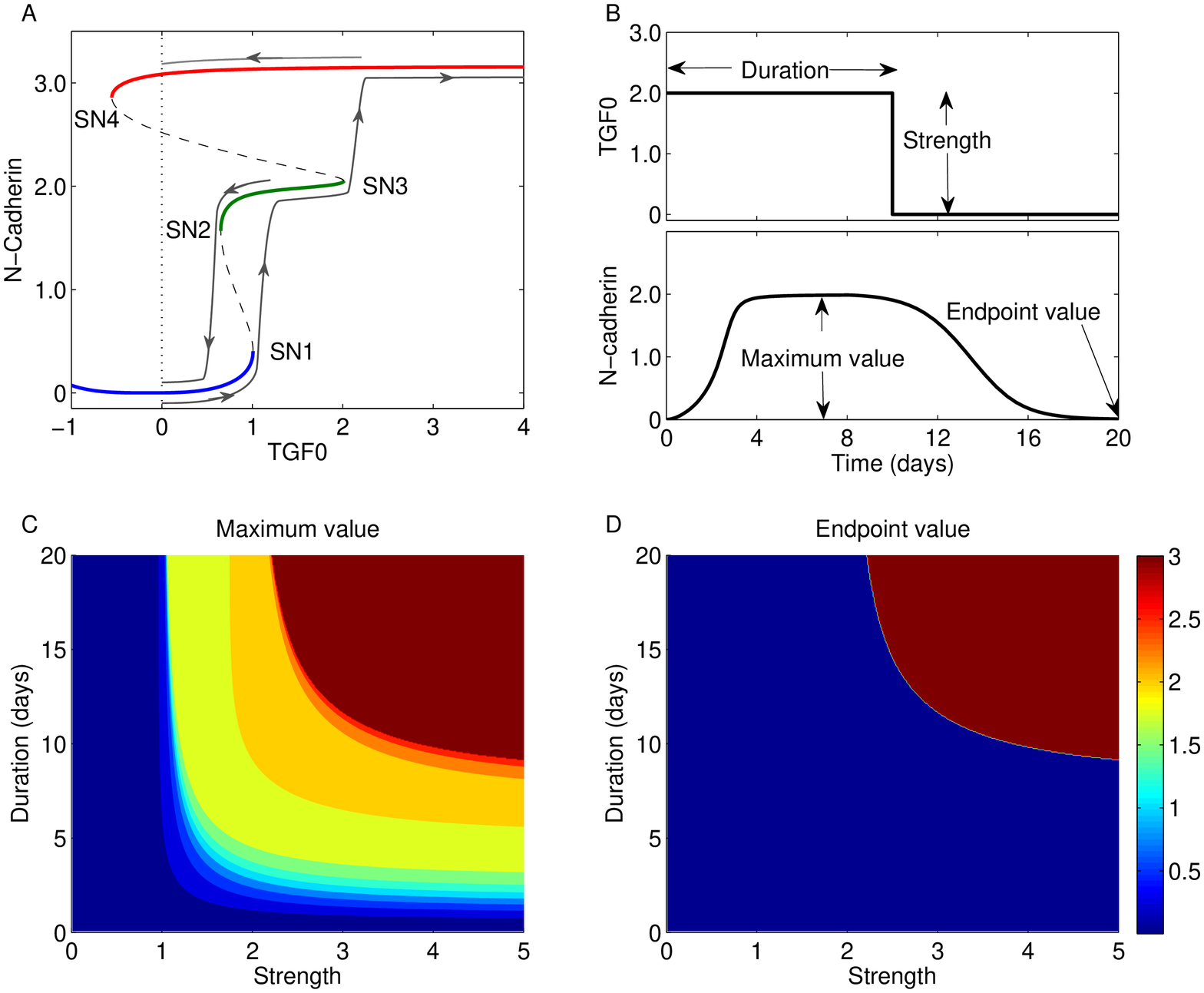}\\
    \caption{}
    \label{fig:three}
  \end{center}
\end{figure}

\clearpage
\begin{figure}
  \begin{center}
    \includegraphics[width=5in]{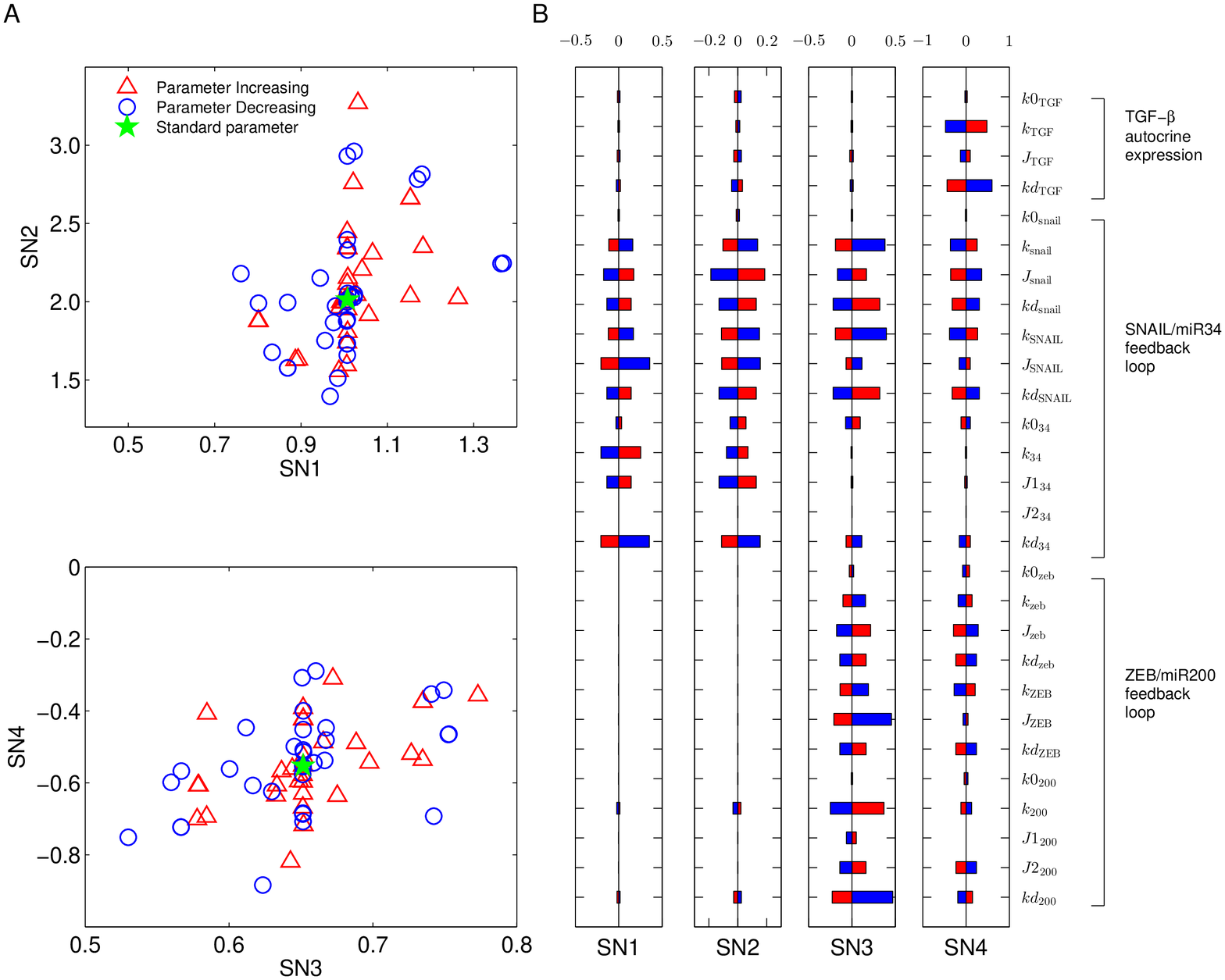}\\
    \caption{}
    \label{fig:four}
  \end{center}
\end{figure}

\clearpage
\begin{figure}
  \begin{center}
    \includegraphics[width=5in]{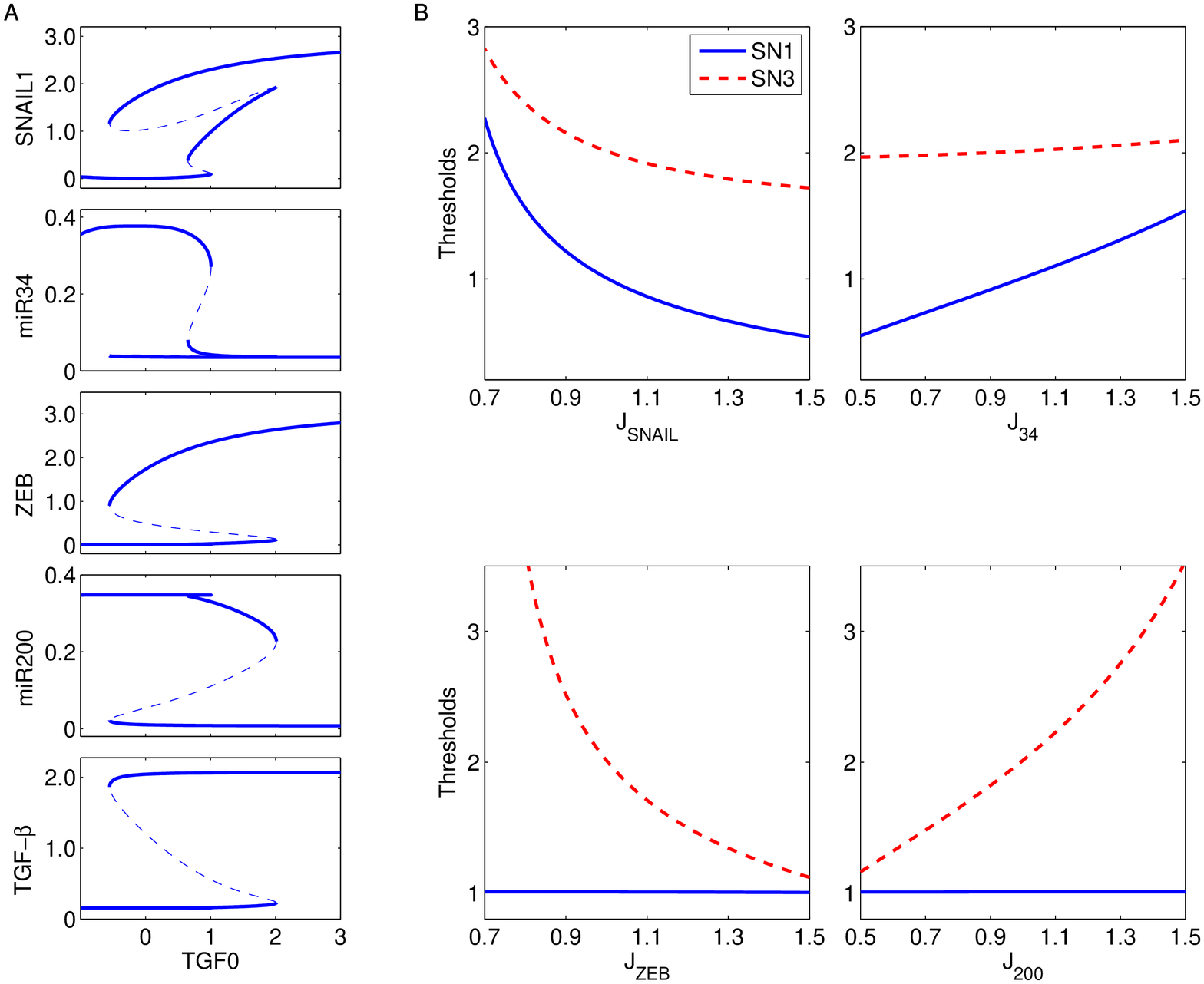}\\
    \caption{}
    \label{fig:five}
  \end{center}
\end{figure}

\clearpage
\begin{figure}
  \begin{center}
    \includegraphics[width=5in]{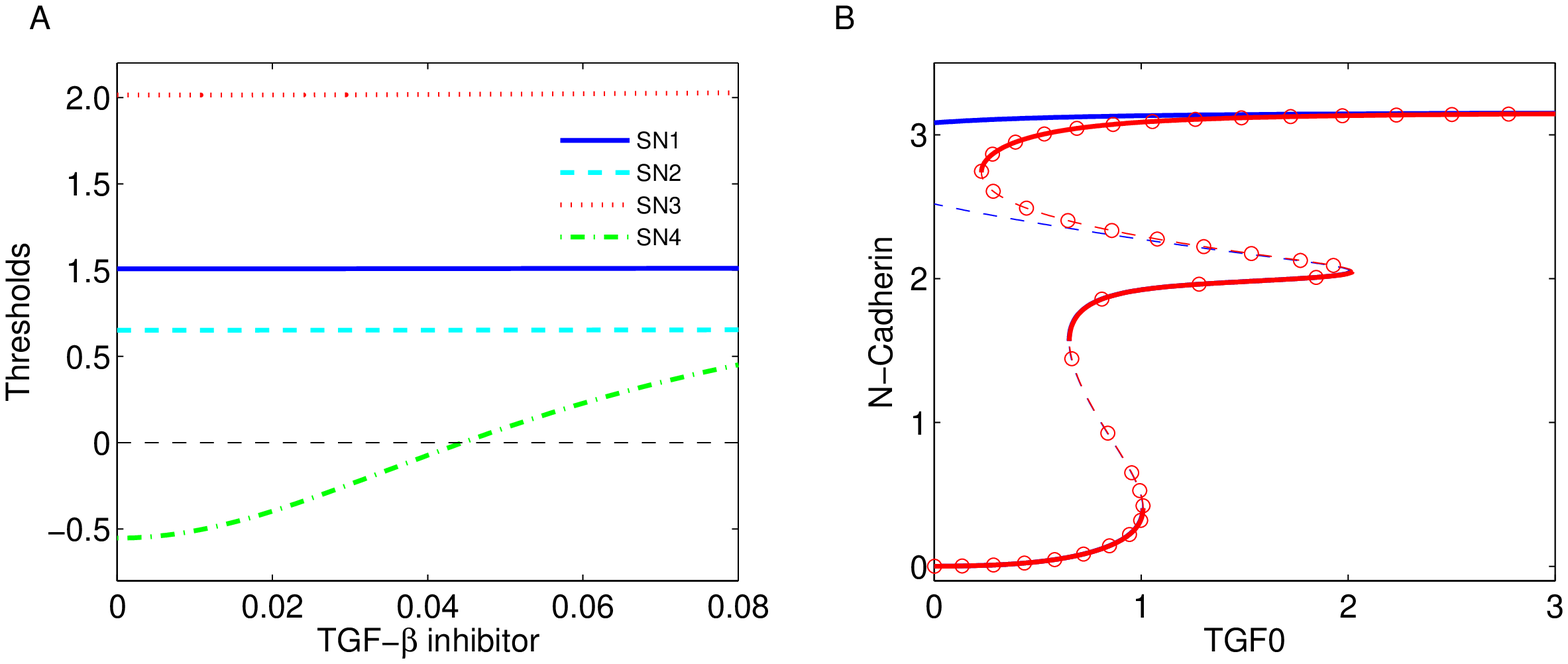}\\
    \caption{}
    \label{fig:six}
  \end{center}
\end{figure}

\clearpage
\begin{figure}
  \begin{center}
    \includegraphics[width=5in]{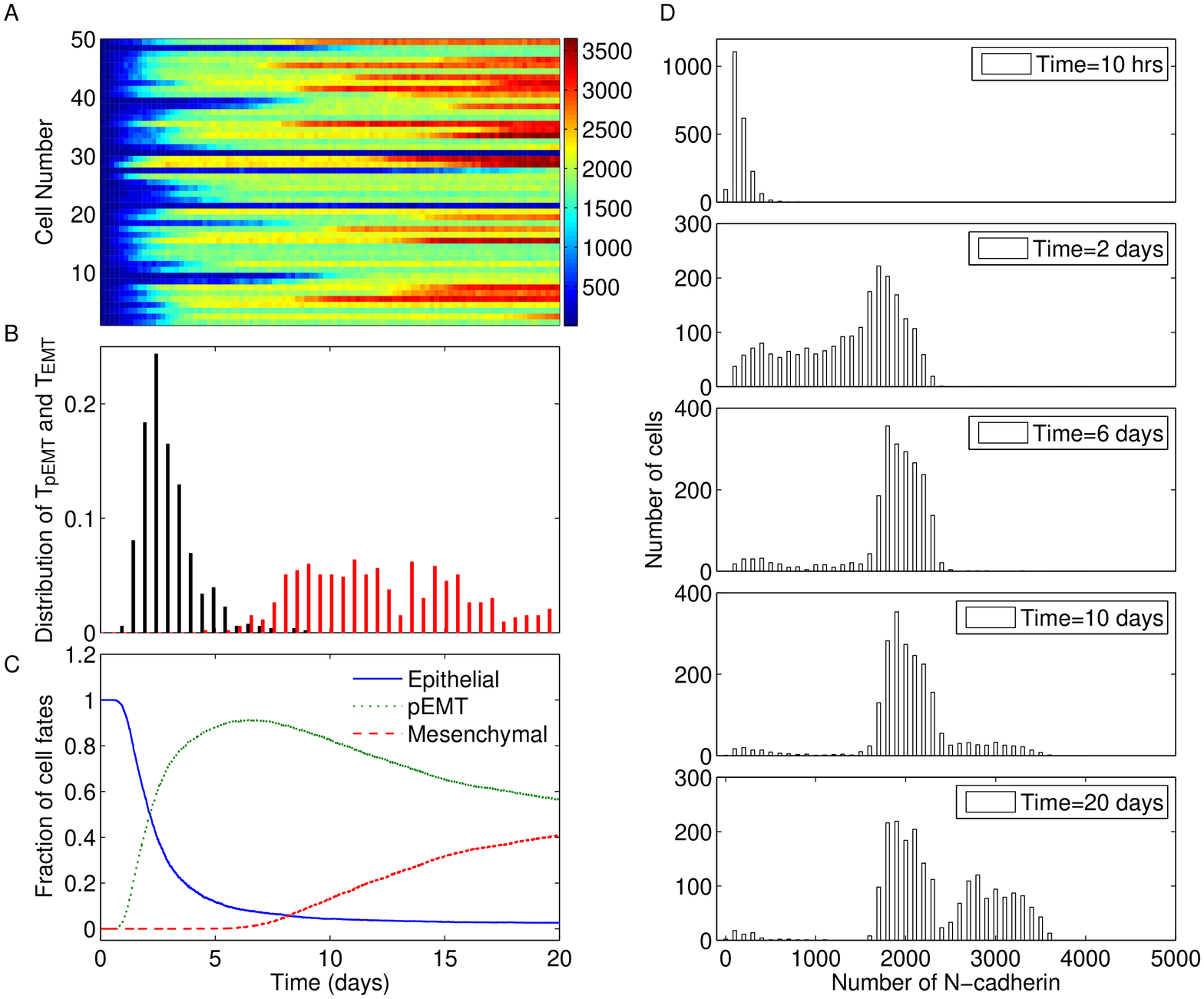}\\
    \caption{}
    \label{fig:seven}
  \end{center}
\end{figure}

\clearpage
\begin{figure}
  \begin{center}
    \includegraphics[width=5in]{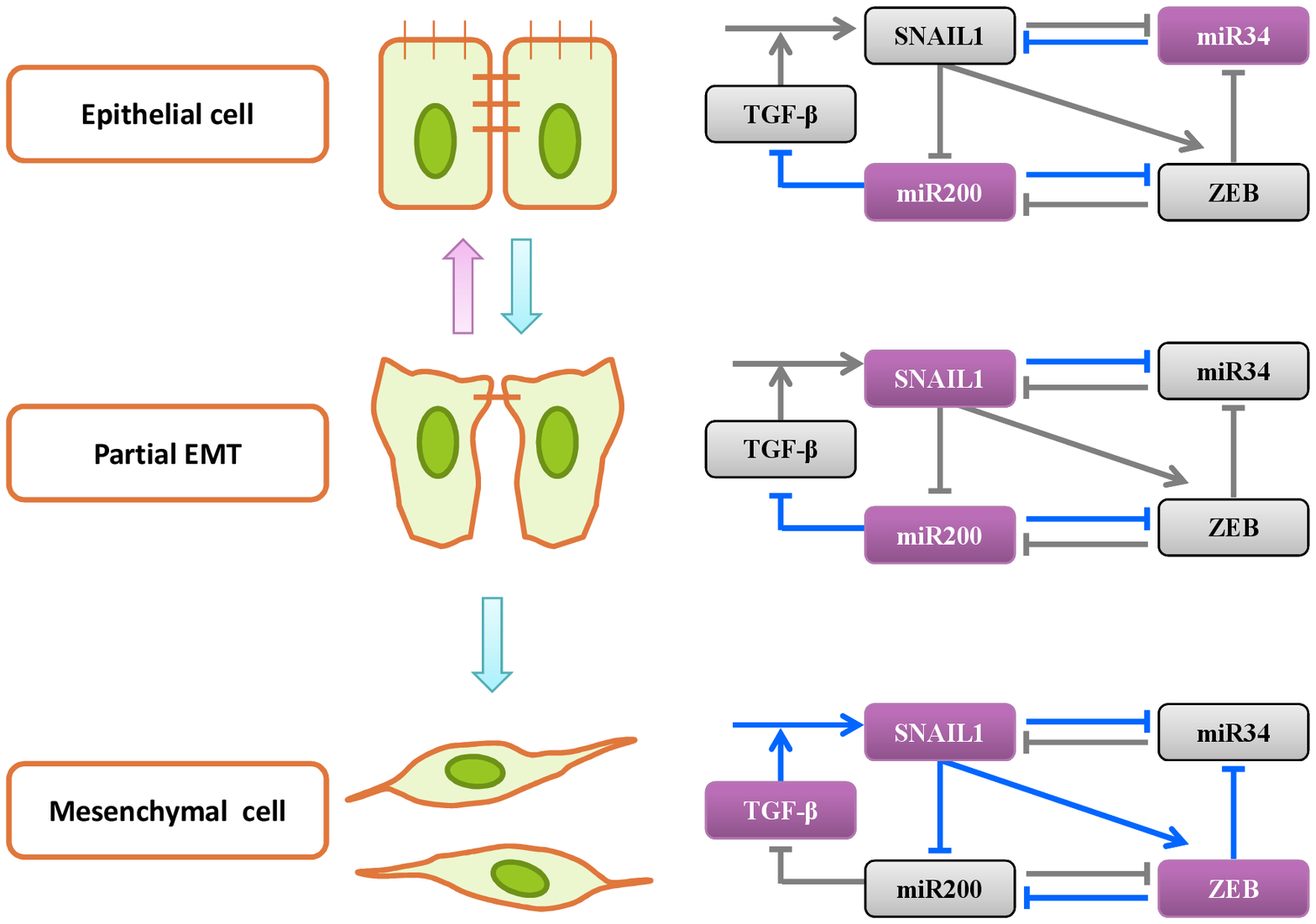}\\
    \caption{}
    \label{fig:eight}
  \end{center}
\end{figure}
\end{document}